\documentclass[%
 reprint,
nofootinbib,
 amsmath,amssymb,
 aps,
]{revtex4-2}

\usepackage{graphicx}
\usepackage{dcolumn}
\usepackage{bm}

\usepackage[colorlinks=true,linkcolor=blue,citecolor=blue,urlcolor=blue]{hyperref}

\newcommand{\conjg}[1]{\ensuremath{\hspace{1pt}\overline{\hspace{-1pt}#1\hspace{-1pt}}}\hspace{1pt}}

        \def\mM{\ensuremath{\mathcal{M}}}

        \def\mP{\ensuremath{\mathcal{P}}}

        \def\mS{\ensuremath{\mathcal{S}}}

\begin{document}

\preprint{APS/123-QED}

\title{Five-body systems with Bethe-Salpeter equations}

\author{Gernot Eichmann$^1$}
\email[]{gernot.eichmann@uni-graz.at}
\author{M.T. Peña$^{2,3}$}
\email{teresa.pena@tecnico.ulisboa.pt}
\author{Raul D. Torres$^{2,3}$}
\email[]{raul.torres@tecnico.ulisboa.pt}
\affiliation{$^1$Institute of Physics, University of Graz, NAWI Graz, Universitätsplatz 5, 8010 Graz, Austria}
\affiliation{$^2$Departamento de Física, Instituto Superior Técnico, Universidade de Lisboa, Av. Rovisco Pais 1, 1049-001 Lisboa, Portugal}
\affiliation{$^3$Laboratório de Instrumentação e Física Experimental de Partículas, Av. Prof. Gama Pinto 2, 1649-003 Lisboa, Portugal}

\date{\today}

\begin{abstract}
  We extend the Bethe-Salpeter formalism to systems made of five valence particles.
  Restricting ourselves to two-body interactions, we derive the subtraction terms necessary to prevent overcounting.  
  We solve the five-body Bethe-Salpeter equation numerically for a system of five scalar particles interacting by a scalar exchange boson.
  To make the calculations tractable, we implement properties of the permutation group $S_5$ and
  construct an approximation based on intermediate two- and three-body poles.
	We extract the five-body ground and excited states along with the spectra obtained from the two-, three-, and four-body equations.
   In the limit of a massless exchange particle, the two-, three, four- and five-body states coexist within a certain range of the coupling strength, 
   whereas for heavier exchange particles the five-body system becomes Borromean.
   Our study serves as a building block for the calculation of pentaquark properties using functional methods.
\end{abstract}

\maketitle

\section{ Introduction}
\label{Sec I. Introduction}

There are presently five pentaquark candidates with minimal quark content $qqqc\overline{c}$ (with $q=u,d,s$), 
which have been observed by the LHCb collaboration in the $J/\Psi p$ and $J/\Psi \Lambda$ invariant mass spectra~\cite{LHCb:2015yax,LHCb:2019kea,LHCb:2022ogu}. The proximity of their peaks to meson-baryon thresholds suggests a molecular explanation in terms of meson-baryon molecules. 
This picture has been frequently employed in effective field theory and model calculations,
 in analogy to several exotic meson candidates in the charmonium sector, see e.g.~\cite{Chen:2016qju,Lebed:2016hpi,Esposito:2016noz,Lebed:2016hpi,Guo:2017jvc,Ali:2017jda,Liu:2019zoy,Brambilla:2019esw}.

In general, it is a highly interesting question how a state made of valence quarks and/or antiquarks 
transforms into a molecular state, given that quantum field theory does not provide the means to rigorously distinguish between these scenarios.
In the analogous case of  four-quark ($qq\bar{q}\bar{q}$) states,
one way to identify such a mechanism is the Bethe-Salpeter equation (BSE)~\cite{Eichmann:2015cra,Eichmann:2020oqt,Hoffer:2024alv}:   
In its solution, the four-quark wave function dynamically develops two-body clusters
in the form of meson-meson and diquark-antidiquark configurations.
For baryons with three valence quarks, the analogue is the formation of internal diquark clusters~\cite{Eichmann:2016yit,Barabanov:2020jvn}.  
In the case of pentaquarks, the internal clusters are mesons and baryons, which naturally leads 
to molecular configurations in the vicinity of meson-baryon thresholds. 

Motivated by these ideas,
in the present work we extend the Bethe-Salpeter formalism~\cite{Salpeter:1951sz,Gell-Mann:1951ooy,Schwinger:1951ex,Dyson:1949ha} to five-body systems. 
As a first application we consider the massive Wick-Cutkosky model~\cite{Wick:1954eu,Cutkosky:1954ru,Nakanishi:1969ph},
which describes the interactions of scalar particles through scalar exchanges. 
The applications of the two-body BSE in this model are well explored by now~\cite{Scarf:1955,Nakanishi:1969ph,Nakanishi:1963zz,Nakanishi:1988hp,Seto:1992rc,Fukui:1995be,
Kusaka:1995za,Nieuwenhuis:1996qx,Kusaka:1997xd,Ahlig:1998qf,Sauli:2001we,Karmanov:2005nv,Frederico:2011ws,Frederico:2013vga,Gutierrez:2016ixt,Eichmann:2019dts,Eichmann:2021vnj,Fornetti:2024uel}  
and the three-body BSE has been investigated in~\cite{Karmanov:2009bhn,Ydrefors:2017nnc,Ydrefors:2019jvu}. 
In the present study we extend this approach to also calculate the spectrum of four- and five-body states.

The paper is organized as follows. In Sec.~\ref{Sec II. Five-body Bethe Salpeter Equation} we establish the five-body BSE and
derive the subtraction terms for the two-body kernel that are necessary to avoid overcounting. We discuss approximations
based on the multiplet structure of the permutation group $S_5$~\cite{Eichmann:2025etg} and the emergence of internal two- and three-body poles.
In Sec.~\ref{Sec III. Results} we present our results, and we conclude in Sec.~\ref{sec:summary}.
Appendix~\ref{sec:nbody} collects technical details on $n$-body BSEs.
We employ a Euclidean metric throughout this work, see~\cite{Eichmann:2016yit} for conventions.

\section{Five-body equation} 
\label{Sec II. Five-body Bethe Salpeter Equation} 

\subsection{General form of the BSE}

\begin{figure}[!t]
  \includegraphics[width=0.9\columnwidth]{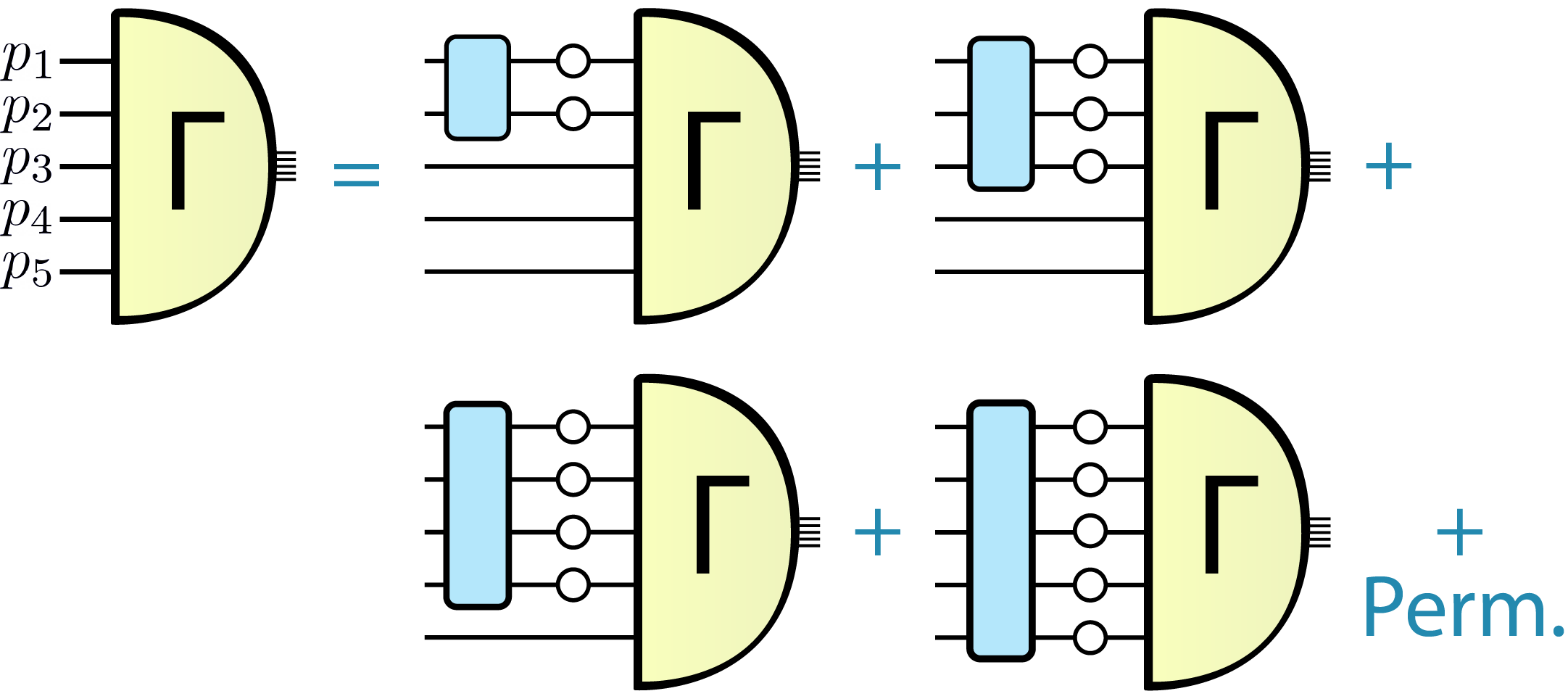}
  \caption{Five-body Bethe Salpeter equation with two-, three-, four- and five-body kernels.}
  \label{Fig: Five Body - All kernels} 
\end{figure}

Our starting point is the homogeneous BSE for a five-body system shown in Fig.~\ref{Fig: Five Body - All kernels}:
\begin{equation} \label{Eq. 5Body BSE}
  \Gamma^{(5)}=K^{(5)}G^{(5)}_{0}\,\Gamma^{(5)}.	
\end{equation}
Here, $K^{(5)}$ is the five-body interaction kernel which consists of two-, three-, four- and five-body interactions,
$G^{(5)}_{0}$ is the product of five dressed particle propagators, and $\Gamma^{(5)}$ is the five-body Bethe-Salpeter amplitude. 
In this compact notation, each multiplication represents an integration over all   four-momenta in the loops.

Like any other homogeneous BSE, the five-body BSE can be derived from the pole behavior of the 5-body scattering matrix $T^{(5)}$, which is a ten-point correlation function
and satisfies the scattering equation 
\begin{equation}   \label{Eq. 5Body Scattering equation}
  T^{(5)}=K^{(5)}+K^{(5)}G^{(5)}_{0}T^{(5)}\,.
\end{equation}
At a given bound-state or resonance pole with mass $M$, it assumes the form
\begin{equation}
   T^{(5)} \longrightarrow \frac{\Gamma^{(5)}\,\conjg{\Gamma}^{(5)}}{P^2+M^2} \,,
\end{equation}
where $\conjg{\Gamma}^{(5)}$ is the charge-conjugate amplitude.
Comparing the residues on both sides of the equation yields the homogeneous equation~\eqref{Eq. 5Body BSE}.

In the following we neglect irreducible three-, four- and five-body forces, so that the resulting kernel consists of irreducible two-body interactions only.
We will denote this two-body kernel by $K$ and assume that $K^{(5)} \approx K$.

\subsection{Subtraction diagrams} \label{sec:subtr}

Like in the case of the four-body equation~\cite{Huang1975,Kvinikhidze1992,Heupel:2012ua},   
a naive summation of two-body kernels leads to overcounting in Eq.~\eqref{Eq. 5Body Scattering equation} and
one needs subtraction terms.
In a five-body system there are ten possible two-body kernels 
\begin{equation}\label{10-top}
  \begin{split}
    K_a \in \big\{ & K_{12}\,, \; K_{13}\,, \; K_{14}\,, \; K_{15}\,, \; K_{23}\,, \\
                   & K_{24}\,, \; K_{25}\,, \; K_{34}\,, \; K_{35}\,, \; K_{45}  \big\} \,,
  \end{split}
\end{equation}
where the indices label the valence particles, and 15 independent double-kernel configurations of the form
\begin{equation}\label{15-top}
  \begin{split}
    K_a\,K_b \in \big\{ & K_{12}\,K_{34}\,, \; K_{12}\,K_{35}\,, \; K_{12}\,K_{45}\,, \; \\
                        & K_{13}\,K_{24}\,, \; K_{13}\,K_{25}\,, \; K_{13}\,K_{45}\,, \; \\
                        & K_{14}\,K_{23}\,, \; K_{14}\,K_{25}\,, \; K_{14}\,K_{35}\,, \; \\
                        & K_{15}\,K_{23}\,, \; K_{15}\,K_{24}\,, \; K_{15}\,K_{34}\,, \; \\
                        & K_{23}\,K_{45}\,, \; K_{24}\,K_{35}\,, \; K_{25}\,K_{34} \big\}\,.
  \end{split}
\end{equation}
By contrast, in a four-body system there are only six two-body kernels and three double-kernel configurations.

\begin{figure}[!t]
	\includegraphics[width=0.9\columnwidth]{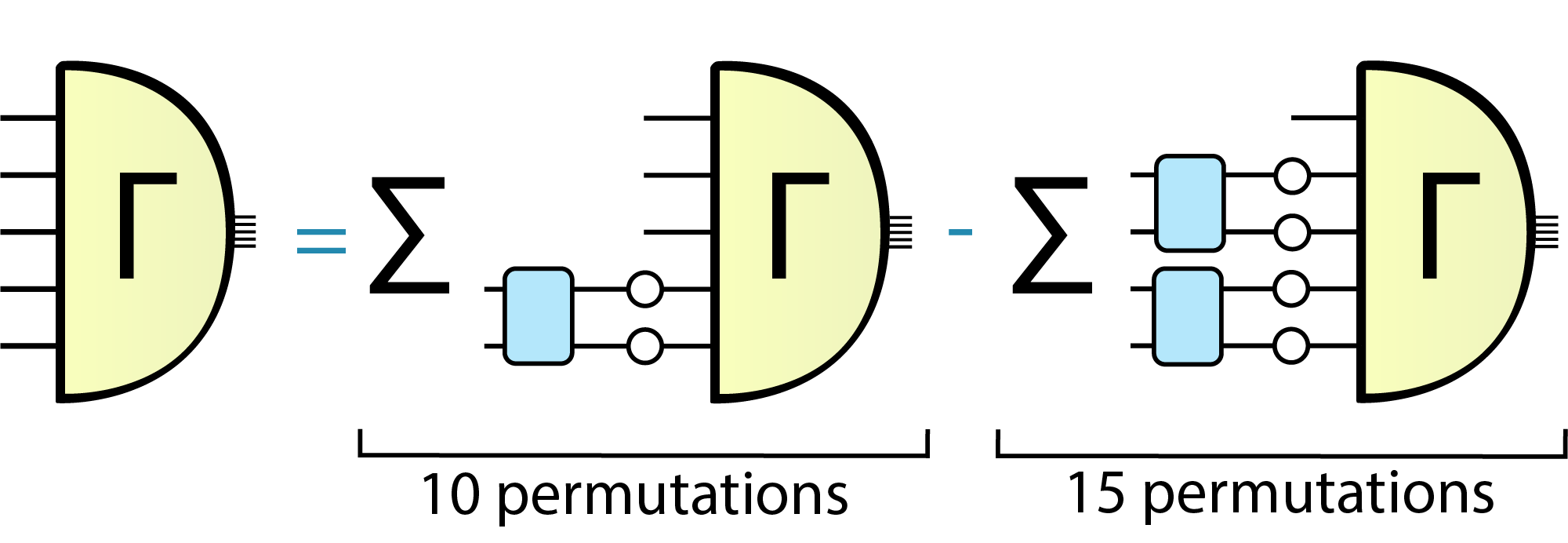}
	\caption{Five-body BSE with two-body kernels and their subtraction terms as given in Eq.~\eqref{full-kernel}.} 
	\label{Fig: FiveBody BSE} 
\end{figure}

If we now define
\begin{equation} \label{K1-K2}
   K_1 := \sum_{a}^{10} K_a\,, \qquad
   K_2 := \sum_{a\neq b}^{15} K_a\,K_b\,,  
\end{equation}
one can show that the combination 
\begin{equation}\label{full-kernel}
   K = K_1 - K_2
\end{equation} 
is free of overcounting, i.e., each possible monomial of the $K_a$  appears exactly once and with coefficient 1 
in the scattering matrix $T^{(5)} = K + K^2 + K^3 + \dots$ that follows from Eq.~\eqref{Eq. 5Body Scattering equation} 
by iteration, where we suppressed the propagator factors $G_0^{(5)}$ for brevity.
The resulting equation is shown in Fig.~\ref{Fig: FiveBody BSE}.

Eq.~\eqref{full-kernel} can  also be derived as follows.
We define the complementary three-body kernel $K_{a'}$ for a given two-body kernel $K_a$ as
\begin{equation}\label{Ka'}
\begin{split}
   a = 12: \quad &K_{a'} = K_{345} = K_{34} + K_{35} + K_{45}\,, \\
   a = 13: \quad &K_{a'} = K_{245} = K_{24} + K_{25} + K_{45}\,, 
\end{split}
\end{equation}
and so on. Now suppose all interactions between the subsystems $a$ and $a'$ (say, $a=12$ and $a'=345$) were switched off. In that case,
the full correlation function $G^{(5)} = G^{(5)}_{0} + G^{(5)}_{0}\,T^{(5)} \,G^{(5)}_{0}$ must factorize into the product $G_a\,G_{a'}$.
Suppressing again the propagators in the notation, we have
\begin{align}
   G^{(5)} &= 1 + T^{(5)} \stackrel{!}{=} G_a\,G_{a'} = (1+T_a)(1+T_{a'}) \\
   &\Rightarrow \; T^{(5)} = T_a + T_{a'} + T_a\,T_{a'}\,.  \label{Taa'}
\end{align} 
Here, $T_a$ and $T_{a'}$ are the scattering matrices for the two- and three-body subsystems,
which satisfy scattering equations analogous to Eq.~\eqref{Eq. 5Body Scattering equation} (written symbolically):
\begin{equation}\label{TaTa'}
\begin{split}
   T_a &= K_a\,(1+T_a) = \frac{K_a}{1-K_a}\,, \\ 
   T_{a'} &= K_{a'}\,(1+T_{a'}) = \frac{K_{a'}}{1-K_{a'}} \,.
\end{split}
\end{equation}
Plugging Eq.~\eqref{TaTa'} into~\eqref{Taa'} yields
\begin{equation}
   T^{(5)} = K\,(1+T^{(5)}) = \frac{K}{1-K}  
\end{equation}
with $K$ given by $K = K_a + K_{a'} - K_{a}\,K_{a'}$.
This is just Eq.~\eqref{full-kernel} 
if the interactions between the clusters (12) and (345) are switched off,
because in that case the kernels $K_1$ and $K_2$ in Eq.~\eqref{K1-K2} reduce to
\begin{equation}
\begin{split}
    K_1 &= K_{12} + K_{34} + K_{35} + K_{45}\,, \\
    K_2 &= K_{12}\,(K_{34} + K_{35} + K_{45})
\end{split}
\end{equation}
and therefore $K_1 - K_2 = K_a + K_{a'} - K_{a}\,K_{a'}$ for $a=12$ and $a'=345$.
Since this relation holds for any combination of clusters $aa'$, Eq.~\eqref{full-kernel}
is the full two-body kernel.

We also note a subtle difference compared to the four-body equation.
In that case, Eq.~\eqref{full-kernel} contains six single-kernel terms and three double-kernel subtraction terms.
This can also be written as the sum over three topologies (12)(34), (13)(24) and (14)(23),
where each contribution is given by $K_a + K_{a'} - K_{a}\,K_{a'}$ (e.g., $K_{12} + K_{34} - K_{12}\,K_{34}$).
In the five-body case this cannot be directly taken over due to the different meaning of $K_a$ and $K_{a'}$.
However, it is still possible to define a kernel $K_{aa'}$ for each single two-body topology in Eq.~\eqref{10-top} such that
\begin{equation}\label{Kaa'}
   K = \sum_{aa'}^{10} K_{aa'} 
\end{equation}
is the sum over the ten topologies.
To this end, we define
\begin{equation} 
    K_3 := \sum_{a'}^{10} K_{a'} \,, \qquad
    K_4 := \sum_{aa'}^{10} K_a\,K_{a'}\,.
\end{equation}
Because each $K_{a'}$ is the sum of three two-body kernels, 
and because $K_1$ and $K_2$ contain 10 and 15 terms, respectively,
this entails $K_3 = 3K_1$ and $K_4 = 2K_2$.
Thus we can write the full two-body kernel $K$ as
\begin{equation}
   K =  K_1 - K_2 = \alpha\,K_1 + (1-\alpha)\,\frac{K_3}{3} - \frac{K_4}{2}  \,,
\end{equation}
where $\alpha$ is an arbitrary parameter. 
Each contribution is now a sum over 10 terms,  
so one can read off the single-topology kernel $K_{aa'}$ in Eq.~\eqref{Kaa'}.
For example, choosing  $\alpha = 1$ yields $K_{aa'} = K_a - K_a\,K_{a'}/2$,  
whereas $\alpha=\frac{1}{4}$ gives
\begin{equation}\label{2-body-kernel}
   K_{aa'} = \frac{K_a}{4}  + \frac{K_{a'}}{4}  - \frac{K_a\,K_{a'}}{2}\,.
\end{equation}

\subsection{Explicit form of the BSE} 

To write down the explicit form of the five-body equation, 
we consider a scalar system made of five scalar particles;
the generalization to particles with spin is straightforward.
The Bethe-Salpeter amplitude $\Gamma(\{p_i\})$ in Fig.~\ref{Fig: Five Body - All kernels} depends on five momenta $p_1 \dots p_5$, 
whose sum is the total onshell momentum $P$ with $P^2 = -M^2$.
According to Eq.~\eqref{full-kernel} and Fig.~\ref{Fig: FiveBody BSE}, the five-body equation can be written as
\begin{equation}\label{5bse-1}
   \Gamma(\{p_i\}) = \sum_{a}^{10} \Gamma_{(a)}(\{p_i\}) - \sum_{a\neq b}^{15} \Gamma_{(a,b)}(\{p_i\})\,,
\end{equation}
where $\Gamma_{(a)}(\{p_i\})$ is the diagram with a two-body kernel attached to the particle pair $a =(a_1, a_2)$ from Eq.~\eqref{10-top},
and $\Gamma_{(a,b)}(\{p_i\})$ is the diagram with two kernels attached to the pairs $a =(a_1, a_2)$ and $b=(b_1,b_2)$ from Eq.~\eqref{15-top}:
\begin{equation}\label{5bse-2}
\begin{split}
   \Gamma_{(a)}(\{p_i\})  &= \int \!\!\frac{d^4r}{(2\pi)^4}\, K(p_{a_1},q_{a_1},p_{a_2},q_{a_2}) \\
                           & \quad \times D(q_{a_1})\,D(q_{a_2})\,\Gamma(\{p_i\},a)\,, \\
   \Gamma_{(a,b)}(\{p_i\}) &= \int \!\!\frac{d^4r}{(2\pi)^4}\,  K(p_{a_1},q_{a_1},p_{a_2},q_{a_2}) \\
                           & \quad \times  D(q_{a_1})\,D(q_{a_2})\,\Gamma_{(b)}(\{p_i\},a)\,.
\end{split}
\end{equation}
Here, $K(p_{a_1},q_{a_1},p_{a_2},q_{a_2})$ are the two-body kernels and $D(p^2)$ the single-particle propagators.
The particle momenta inside the loop, where $r$ is the exchanged four-momentum, 
are given by
\begin{equation}\label{exch-mom}
   q_{a_1} = p_{a_1} - r\,, \quad q_{a_2} = p_{a_2} + r\,.
\end{equation}
The amplitudes inside the loop are
\begin{equation}
   \Gamma(\{p_i\},a) = \Gamma(p_1, \dots, q_{a_1}, \dots q_{a_2}, \dots p_5)\,,
\end{equation}
where $p_{a_1}$ is replaced by $q_{a_1}$ and $p_{a_2}$ by $q_{a_2}$.

In practice it is useful to work with the total momentum $P$ and four relative momenta $q$, $p$, $k$, $l$ instead
of the five particle momenta $p_i$ with $i=1\dots 5$. 
To this end, we employ the momenta
\begin{equation}
  \begin{split}
    q &= \frac{p_1-p_5}{2}\,,\qquad  p = \frac{p_2-p_5}{2}\,, \\ 
    k &= \frac{p_3-p_5}{2}\,,\qquad  l = \frac{p_4-p_5}{2}\,. 
  \end{split}
\end{equation}
The amplitude $\Gamma(q,p,k,l,P)$ then depends on 15 Lorentz invariants that 
can be formed from these momenta, namely
\begin{equation}\label{li-variables} \renewcommand{\arraystretch}{1.4}
  \begin{array}{l}
    q^2\,, \\
    p^2\,, \\
    k^2\,, \\
    l^2\,, \\        
    P^2\,, 
  \end{array}  \qquad
	\begin{array}{rl}
    \omega_1 &\!\!= q\cdot p\,, \\
    \omega_2 &\!\!= q\cdot k\,, \\
    \omega_3 &\!\!= q\cdot l\,, \\
    \omega_4 &\!\!= p\cdot k\,, \\        
    \omega_5 &\!\!= p\cdot l\,, \\
    \omega_6 &\!\!= k\cdot l\,, 
  \end{array}  \qquad
  \begin{array}{rl}
    \eta_1 &\!\!= q\cdot P\,, \\
    \eta_2 &\!\!= p\cdot P\,, \\
    \eta_3 &\!\!= k\cdot P\,, \\
    \eta_4 &\!\!= l\cdot P\,.
  \end{array}  
\end{equation}   
These can be arranged into multiplets of the permutation group $S_5$,
namely two singlets, two quartets and a quintet~\cite{Eichmann:2025etg}. The singlet variables are
\begin{equation}\label{S0}
  \mS_0 = \frac{1}{5}\left(  q^2 + p^2 + k^2 + l^2 - \frac{1}{2}\sum_{i=1}^6 \omega_i \right) 
\end{equation}
and $P^2 = -M^2$.
One quartet is constructed from the angular variables $\eta_i$, and the remaining variables are distributed over a quartet and a quintet.
In four spacetime dimensions there is an additional relation between these 15 variables 
so that only 14 are independent. In practice this leads to a nontrivial relation
between the quintets, quartets and singlets which involves five powers in the variables~\eqref{li-variables},
see Appendix D of Ref.~\cite{Eichmann:2025etg} for details.

Table~\ref{tab:variables} shows the extension of the multiplet construction to general $n$-body systems, which are subject to the permutation group $S_n$~\cite{Eichmann:2025etg}.
For any $n$ one can construct a singlet $\mS_0$ analogous to Eq.~\eqref{S0}, and $P^2$ is always a singlet.
A general $n$-body system has $n-1$ relative momenta  and hence $n-1$ angular variables $\eta_i$, which form an $(n-1)$-dimensional multiplet of $S_n$.
The variables $p_1^2 \dots p_n^2$ form another $(n-1)$-plet, with their sum being constrained by $\mS_0$.
In total there are $n(n+1)/2$ Lorentz invariants, so the difference $n(n-3)/2$ gives another multiplet (denoted by $\mM$ in Table~\ref{tab:variables}).
For example, a two-body system forms two singlets ($q^2$ and $P^2$) and an antisinglet ($q\cdot P$).
A three-body system gives two singlets ($\mS_0$ and $P^2$) and two doublets~\cite{Eichmann:2014xya,Eichmann:2011vu}, 
and a four-body system two singlets ($\mS_0$ and $P^2$), a doublet and two triplets~\cite{Eichmann:2015nra}.
For five-body systems one also encounters for the first time the dimensional constraint relating the Lorentz invariants, 
because $n$ four-vectors can only depend on $4n-6$ independent variables.

While a system depending on such a large number of variables is extremely costly to solve numerically,
the $S_n$ construction allows one to switch off entire multiplets without                                        
affecting the symmetries of the system. This is especially useful for constructing approximations   
where one singles out the multiplets with the largest impact on the dynamics.
Previous solutions of two-, three- and four-body systems show that the dependence on the angular variables $\eta_i$
is usually small or even negligible~\cite{Eichmann:2015cra}.
Similarly, Bose-symmetric $n$-point functions like the three- and four-gluon vertex,
which are obtained from Table~\ref{tab:variables} by setting $P^2=0$ and all $\eta_i=0$,
show a planar degeneracy and depend mainly on $\mS_0$~\cite{Eichmann:2014xya,Pinto-Gomez:2022brg,Ferreira:2023fva,Aguilar:2023qqd,Aguilar:2024fen}.
   
   \begin{table}
  \centering
  \begin{tabular}{c @{\quad}  c @{\;\;} c @{\;\;} c @{\;\;} c @{\;\;} c @{\quad} c @{\quad} c } \hline\noalign{\smallskip}
    $n$   & $\mS_0$ & $P^2$ & $\eta_i$ & $p_1^2 \dots p_n^2$ & $\mM$  & Total         & Indep.    \\ \noalign{\smallskip}\hline\noalign{\smallskip}
    $2$        & 1       & 1     & 1        & $-$           & $-$     & 3             & 3         \\[0.5mm]
    $3$        & 1       & 1     & 2        & 2             & $-$     & 6             & 6         \\[0.5mm]
    $4$        & 1       & 1     & 3        & 3             & $2$     & 10            & 10        \\[0.5mm]
    $5$        & 1       & 1     & 4        & 4             & $5$     & 15            & 14        \\[0.5mm]
    $6$        & 1       & 1     & 5        & 5             & $9$     & 21            & 18        \\ \noalign{\smallskip}\hline\noalign{\smallskip}
    $n$        & 1       & 1     & $n-1$    & $n-1$         & $n(n-3)/2$ & $n(n+1)/2$    & $4n-6$    \\\noalign{\smallskip}\hline
  \end{tabular}
  \caption{$S_n$ multiplet counting for an $n$-body system~\cite{Eichmann:2025etg}, see text for the discussion.}
  \label{tab:variables}
\end{table}

The crucial observation from four-body systems is that the BSE dynamically generates intermediate two-body poles
in the solution process. Specifically, a four-quark ($qq\bar{q}\bar{q}$) equation in QCD produces intermediate
meson (and diquark) poles and thus dynamically creates resonance channels. In the four-body system these poles
only appear in the doublet $\mM$, so that the dynamics is largely determined by $\mS_0$ and $\mM$.
In a five-body system, on the other hand, there are 10 possible two- and three-body configurations $aa'=(12)(345)$, $(13)(245), \dots$ which are distributed
over the quartet and quintet. Together with $\mS_0$, one would thus still need to include 10 variables in the BSE to capture the important dynamics.
   
Given that the leading momentum dependence of the amplitude beyond the singlet variable $S_0$ comes from the two-body and three-body clusters,
here we follow a more efficient strategy that has been developed in the four-body case~\cite{Wallbott:2019dng,Wallbott:2020jzh,Hoffer:2024alv}: 
We reduce the momentum dependence to $\mS_0$ but include the two- and three-body poles explicitly.
The resulting amplitude then reads 
\begin{equation}
  \Gamma(q,p,k,l,P)\approx f(S_0)\sum_{aa'} \mP_{aa'} \,,
  \label{Eq. 5Body - Pole Ansatz}
\end{equation}
where e.g. for $aa' = (12)(345)$ the two- and three-body poles of the amplitude are given by 
\begin{equation*}
  \mP_{(12)(345)}=\frac{1}{(p_1+p_2)^2+M^{2}_M}\frac{1}{(p_3+p_4+p_5)^2+M^2_B}\,.
	\label{Eq. Pole Ansatz}
\end{equation*}
Here, $M_M$ and $M_B$ are the masses of the two- and three-body subsystems (`mesons' and `baryons'), respectively. 
When plugging this ansatz into the five-body equation (\ref{5bse-1}--\ref{5bse-2}),
the resulting dressing function $f(\mS_0)$  depends only on $\mS_0$.
In this way, the pole ansatz effectively captures the dependence on the remaining variables which is dominated by these poles.

In turn, this procedure requires knowledge of the bound-state masses $M_M$, $M_B$ of the two- and three-body equations
in the same approach. We solve these equations in a sequence by
employing tree-level propagators for scalar constituent particles with mass $m$ and a ladder approximation for a boson exchange with mass $\mu$:
\begin{equation}\label{Eq. 5Body - Propagator and Kernel}
	D(p)=\frac{1}{p^2+m^2}, \quad
    K(p_{a_1},q_{a_1},p_{a_2},q_{a_2})=\frac{g^2}{r^2+\mu^2}\,,
\end{equation}
where $r$ is the exchange momentum according to Eq.~\eqref{exch-mom}. 
We consider equal constituent masses for simplicity, but the generalization to unequal-mass systems is straightforward.
In the following we employ a dimensionless coupling constant
$c$ and mass ratio $\beta$ via
\begin{equation}
  c=\frac{g^2}{(4\pi m)^2}, \qquad \beta=\frac{\mu}{m}\,,
  \label{Eq. Multi-body coupling}
\end{equation}
so that all results only
depend on $c$ and $\beta$ while the mass $m$ drops out.

The details on the \mbox{two-,} three-, four- and five-body equations for the scalar theory are provided in App.~\ref{sec:nbody}.
In the following we distinguish three  approximations when solving these $n$-body equations.
A `full solution' refers to solving the respective BSE without any further approximations on the kinematics in the amplitude.
At present, this is  numerically only feasible for the two- and three body equations.
The `singlet $\times$ pole' approximation refers to Eq.~\eqref{Eq. 5Body - Pole Ansatz}, 
with explicit two- and three-body poles for the five-body equation (`mesons' and `baryons'),
and two-body poles for the three- and four-body equations (`mesons' or `diquarks').
Finally, the `singlet approximation' refers to Eq.~\eqref{Eq. 5Body - Pole Ansatz} without a pole ansatz, i.e., $\Gamma(q,p,k,l,P)\approx f(S_0)$,
which  will be used for comparisons.

\begin{figure}[t]
   \includegraphics[width=1\columnwidth]{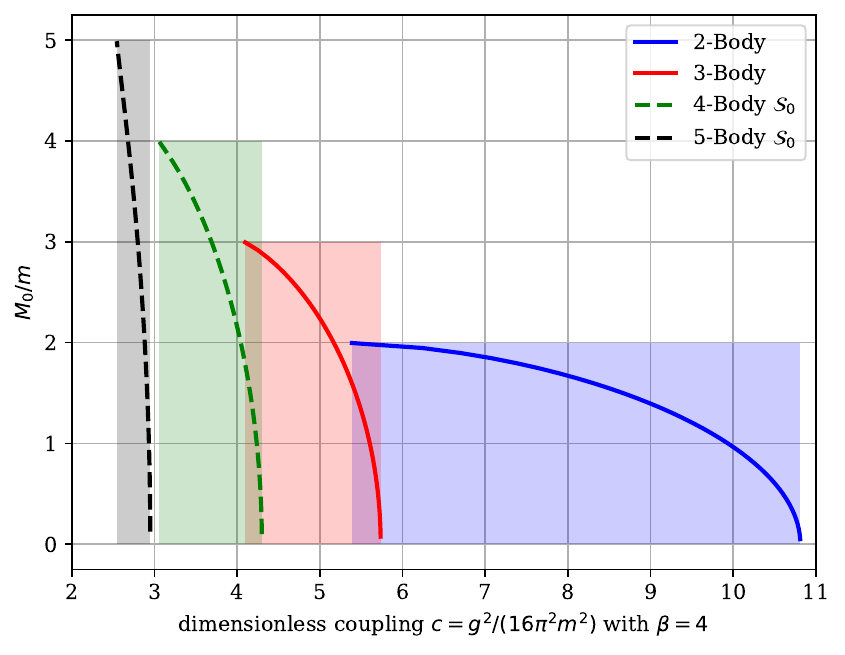} 
   \caption{Ground-state masses obtained from the $n$-body BSEs for $\beta=4$ and different values of the coupling $c$.
     The two- and three-body results (solid curves) are full solutions,
     while the four-and five-body results (dashed curves) are obtained in the singlet approximation.
     The  squares show the regions where ground-state solutions are possible. } 
   \label{Fig: Eigenvalue 4} 
\end{figure}

\section{ Results}
\label{Sec III. Results}

In the following we present our solutions of the two-, three-, four- and five-body equations
in the setup described above. 
In practice the BSEs turn into eigenvalue equations of the form 
\begin{equation}
   \lambda_i(P^2) \,\Psi_i(P^2) =  \mathcal{K}(P^2)\,\Psi_i(P^2)\,,
\end{equation}
where $P^2\in\mathbb{C}$ is the total five-body momentum 
squared, $\mathcal{K}(P^2)$ is the kernel and the $\Psi_i(P^2)$ are its eigenvectors with eigenvalues $\lambda_i(P^2)$ 
for the  ground  ($i=0$) and excited states ($i >0$). 
If the condition $\lambda_i(P^2)=1$ is satisfied, this corresponds to a pole in the scattering matrix at $P^2=-M^2_i$ and determines the respective mass $M_i$.
All results depend on two parameters, the coupling strength $c$ and the mass ratio $\beta$.
We cross-checked our results with the literature; our two-body solutions agree with those obtained in Refs.~\cite{Ahlig:1998qf,Eichmann:2019dts}
and our three-body solutions with those in Ref.~\cite{Karmanov:2009bhn}.

Fig.~\ref{Fig: Eigenvalue 4} shows the variation of the ground-state masses $M_0$ 
obtained from the two-, \mbox{three-,} four- and five-body equations  
with the coupling strength $c$  at a fixed value $\beta=4$. 
One can see that for each system a ground state only exists within a certain range of the coupling.
If the binding is too weak, the mass exceeds the respective threshold, and the bound state
will turn into a resonance or virtual state on the second Riemann sheet. If the binding is too strong, 
the squared mass $M_0^2$ becomes negative and the bound state  turns into a tachyon
(see~\cite{Eichmann:2019dts} for explicit examples).
The latter property is presumably an artifact of the ladder approximation:
Because the propagators remain at tree level and the three-point interaction vertices are constant,
the coupling strength $c$ only enters as an overall factor on each ladder kernel.
In more advanced truncations
where the $n$-point functions in the kernel are solved from their Dyson-Schwinger equations,
the masses $M_i(c)$ would eventually approach constant values; see~\cite{Eichmann:2023tjk} for a corresponding study of the scalar two-body equation.

\begin{figure}[t]
  \includegraphics[width=1\columnwidth]{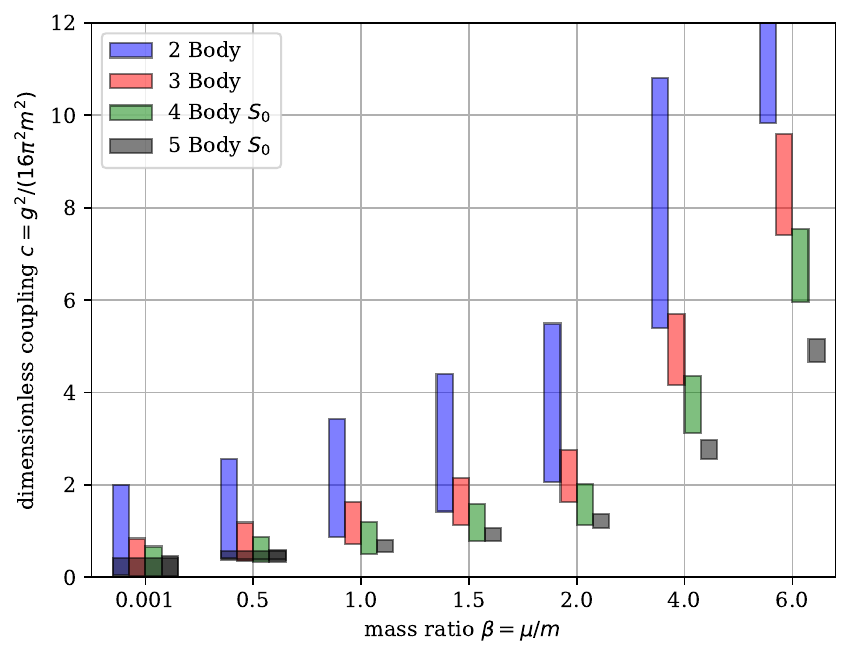} 
  \caption{Coupling ranges where $n$-body ground states are possible shown for different values of $\beta$.
  The horizontal displacement is for better readability.
  The color coding is the same as in Fig.~\ref{Fig: Eigenvalue 4}. The black rectangles for small $\beta$ values are the regions where all solutions coexist.} 
  \label{Fig: Coupling-Bound States} 
\end{figure}

The lower coupling limit, below which the states become unbound, 
implies that below certain values of $c$ not all ground states can coexist.
For example, in Fig.~\ref{Fig: Eigenvalue 4} the three-body equation displays a Borromean behavior for  $c \lesssim 5.5$, 
i.e., it admits a ground state
while there is no corresponding two-body ground state.
Similarly, for $c \lesssim 3$ the five-body equation admits a ground state while there
are no \mbox{two-,} three- or four-body ground states. 
Fig.~\ref{Fig: Coupling-Bound States} displays the resulting coupling ranges for varying values of $\beta$.
The coexistence regions are shown by the black rectangles and only exist
for small $\beta$ values ($\beta \lesssim 0.5$).
For higher values of $\beta$ the five-body system becomes Borromean,
i.e., there is a five-body ground state without corresponding two- and three-body ground states.
 
\begin{figure}[t]
   \includegraphics[width=1\columnwidth]{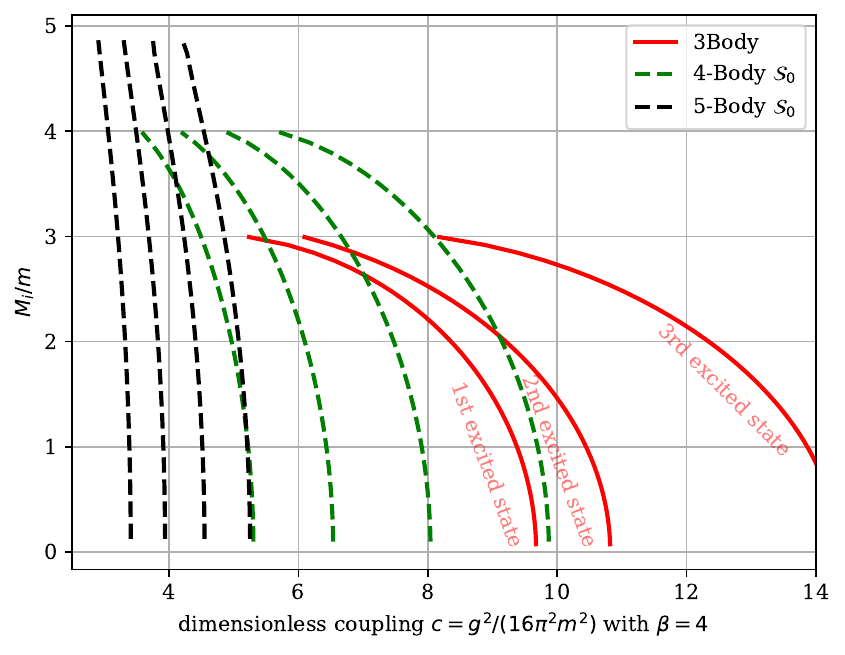} 
   \caption{Radially excited state masses $M_{i>0}$ for $\beta=4$ and different values of the coupling $c$.
   The three-body results (solid curves) are full solutions, 
   while the four- and five-body results (dashed curves) are obtained in the singlet approximation.} 
   \label{Fig: EigenvalueExcited 4} 
\end{figure}

In Fig. \ref{Fig: EigenvalueExcited 4} we also show the radially excited states $M_{i>0}$
from the three-, four- and five-body solutions. For the extraction of the excited states
we use an implementation of the Arnoldi algorithm~\cite{Arpack}.

As a  consequence of the Borromean behavior, 
the five-body equation can dynamically generate two- and three-body ground-state poles only in the coexistence region.
Outside  this region, these poles would  correspond to resonances or virtual states. 
Thus, the pole ansatz~\eqref{Eq. 5Body - Pole Ansatz} can  also only be sensibly applied in the coexistence region,
which is why in Figs.~\ref{Fig: Eigenvalue 4}, \ref{Fig: Coupling-Bound States} and~\ref{Fig: EigenvalueExcited 4} we used 
the singlet approximation for the four- and five-body equations.

\begin{figure}[t]
  \includegraphics[width=1\columnwidth]{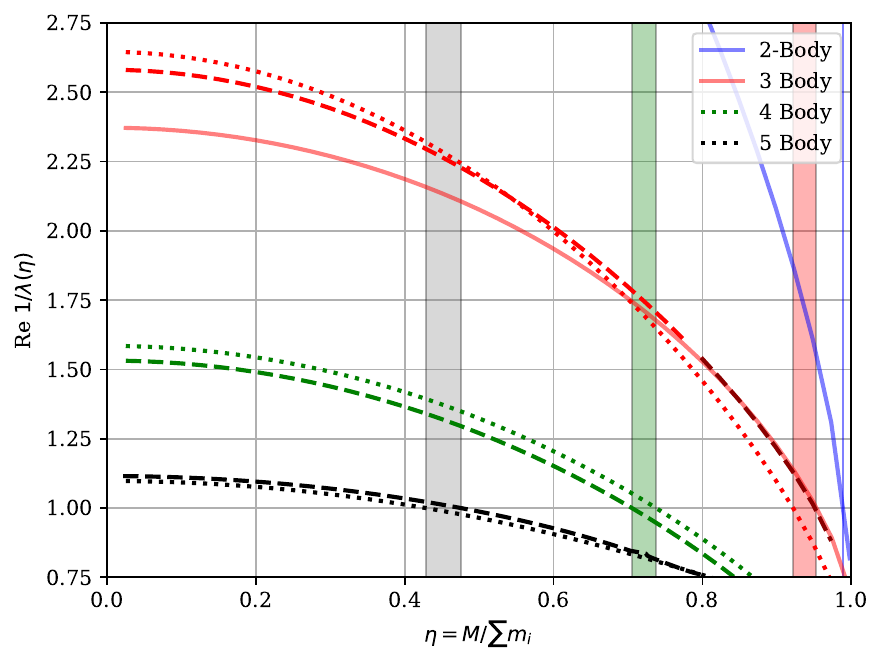} 
  \caption{Eigenvalues of the $n$-body BSEs for $\beta=c=0.5$. 
  Solid curves correspond to full solutions, dotted curves to the singlet approximation 
  and dashed curves to the singlet $\times$ pole approximation.
  The vertical bands show the resulting mass ranges when going from the singlet to the singlet $\times$ pole approximation.} 
  \label{Fig: Eigenvalue Pole} 
  \vspace{-1mm}
\end{figure}

To go beyond this approximation, we must  choose values of $\beta$ and $c$ inside the coexistence region.
This is done in Fig.~\ref{Fig: Eigenvalue Pole} for $\beta = c=0.5$.
Here we plot the inverse ground-state eigenvalues $1/\lambda_0(P^2)$ of the two-, three-, four- and five-body BSEs
as a function of $M$ (corresponding to $P^2 = -M^2$). We divide the mass by the sum of the propagator masses, such that the threshold
of each equation is  $\eta := M/\sum_i m_i = 1$. The  ground-state masses $M_0$ are then obtained from the intersections $\lambda_0(P^2) = 1$.
As before, the solid curves are the results from the full solutions (which are  available for the two- and three-body BSEs).
The dotted curves now correspond to the singlet approximation and the dashed curves  to the singlet $\times$ pole approximation.
From the three-body curves one can see that the singlet $\times$ pole approximation
nicely agrees with the full solution, as long as the mass is not too far from the threshold, 
while the mass obtained with the singlet approximation is  lower.
The singlet $\times$ pole ansatz is therefore a very good approximation of the dynamics in the system. 
The analogous observation in QCD is diquark clustering:
The solution of the three-body Faddeev equation dynamically generates diquark poles, which dominate the behavior of the system,
and the spectra and form factors in the quark-diquark approach
agree well with those of the three-body solution~\cite{Eichmann:2016yit,Barabanov:2020jvn}. Likewise, four-quark ($qq\bar{q}\bar{q}$) systems dynamically
generate meson and diquark poles which dominate their properties~\cite{Eichmann:2015cra,Eichmann:2020oqt,Hoffer:2024alv}.

The vertical bands in Fig.~\ref{Fig: Eigenvalue Pole} show the intersections of the eigenvalue curves with 1
for the singlet and singlet $\times$ pole approximations. The resulting masses differ by $\lesssim 10\%$, so that
already the singlet approximation yields reasonable estimates for them.
However, this  observation does not translate well to QCD; e.g., for light scalar $qq\bar{q}\bar{q}$ systems which are dominated by $\pi\pi$ channels,
the singlet and singlet $\times$ pole approximations lead to very different results due to the small pion mass~\cite{Eichmann:2015cra}.

Because the five-body equation dynamically generates two- and three-body poles,
which is made explicit by the singlet $\times$ pole approximation,
this also lowers the threshold of the system from $M=5m$ to $M=M_M+M_B$, where $M_M$ is the `meson' and $M_B$ the `baryon' mass.
For the parameter values in Fig.~\ref{Fig: Eigenvalue Pole} this leads to the restriction $\eta \lesssim 0.8$.
Above this value, the five-body ground state turns into a resonance or virtual state, and one would need to employ contour deformations 
and analytic continuations to extract its mass~\cite{Eichmann:2019dts,Santowsky:2020pwd}. 
Likewise, in the four-body system the threshold changes from $M=4m$ to $M=2M_M$, and in the three-body system it changes from from $M=3m$ to $M=m+M_M$.

\begin{figure}[t]
  \includegraphics[width=1\columnwidth]{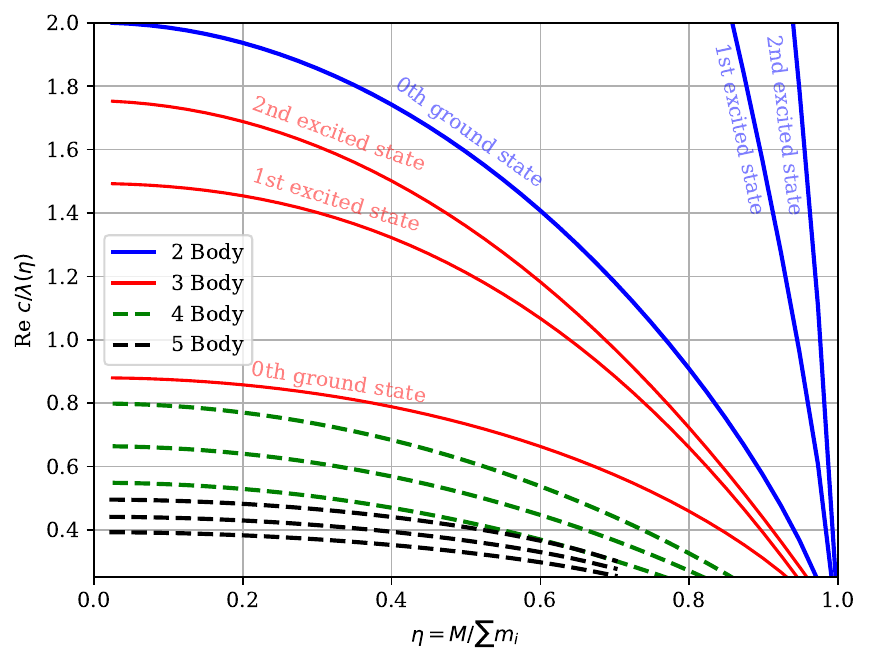} 
  \caption{BSE eigenvalues for $\beta=0.001$, with $c=1/4$ to ensure coexisting solutions for the two-, three-, four- and five-body BSEs (see Fig.~\ref{Fig: Coupling-Bound States}).
    We rescaled the $y$ axis again with $c$ so that the condition $1/\lambda_i = 1$ for the physical solutions
    becomes  $c/\lambda_i = 1/4$,
    which is the baseline in the plot.
   The two- and three-body results (solid curves) are full solutions,
   while the four- and five-body results (dashed curves) are obtained in the singlet $\times$ pole approximation.
   } 
  \label{Fig: Coupling-Mass} 
\end{figure}

Finally, in Fig.~\ref{Fig: Coupling-Mass} we show the analogous eigenvalue plot for $\beta \approx 0$ corresponding
to the massless Wick-Cutkosky model. With our definition~\eqref{Eq. Multi-body coupling} of the coupling strength, 
the inverse eigenvalues of the two-body system for $c=1$ and $M=0$ become integers, which can also be determined analytically~\cite{Cutkosky:1954ru,Ahlig:1998qf}.
For the three-, four- and five-body systems, on the other hand, this does not appear to be the case.

\section{Summary}\label{sec:summary}

We developed the five-body Bethe-Salpeter formalism and
solved the five-body equation for a scalar model in a ladder truncation.
The five-body Bethe-Salpeter amplitude depends  on 14 momentum variables, which can be arranged
in multiplets of the permutation group $S_5$.
To reduce this large number of variables, we employed an approximation in terms of
two- and three-body poles,  which the full amplitude would generate dynamically.
Since this requires knowledge of the two- and three-body bound state masses, 
we also solved the corresponding two-, three- and four-body equations in the same approach.
The two- and three-body equations can be solved without any approximations on the amplitude,
and we find that a pole approximation in the three-body sector works very well.
The approach developed in this work can be extended to QCD in view of  investigating pentaquarks, and work in this direction is underway.

\begin{acknowledgments}
\noindent 
We are grateful to Eduardo Ferreira and Joshua Hoffer for helpful discussions.
This work was supported by the Portuguese Science fund FCT
under grant numbers CERN/FIS-PAR/0023/2021 and PRT/BD/152265/2021 
and by the Austrian Science Fund FWF under grant number 10.55776/PAT2089624.
This work contributes to the aims of the USDOE ExoHad Topical Collaboration, contract DE-SC0023598.
\end{acknowledgments}


\appendix

\section{n-body equations} \label{sec:nbody} 

In this appendix we provide details on the scalar \mbox{three-,} four- and five-body BSEs.
An $n$-body bound state with total four-momentum $P$ is the solution of the $n$-body BSE:
\begin{equation}
   \Gamma^{(n)}=K^{(n)}G^{(n)}_{0}\Gamma^{(n)}\,.
	 \label{Eq. Bethe Salpeter Equation}
\end{equation}
We restrict ourselves to two-body interactions, so that $K^{(n)}$ is the sum of all two-body kernels (possibly including subtraction terms)
and $G^{(n)}_{0}$ the product of the single-particle propagators.
Using the one-boson exchange kernel in Eq.~\eqref{Eq. 5Body - Propagator and Kernel}, the r.h.s.
of each BSE can be written as the sum over the two-body interaction diagrams,
where each contains a four-momentum integration over the exchange-boson momentum $r$, cf.~Eq.~\eqref{exch-mom}.
The explicit solution method for the two-body case can be found in Ref.~\cite{Eichmann:2019dts};
in the following we discuss the three-, four- and five-body BSEs.

\subsection{Three-body system} \label{App. Three-Body System}

\begin{figure}[!t]
   \includegraphics[width=1\columnwidth]{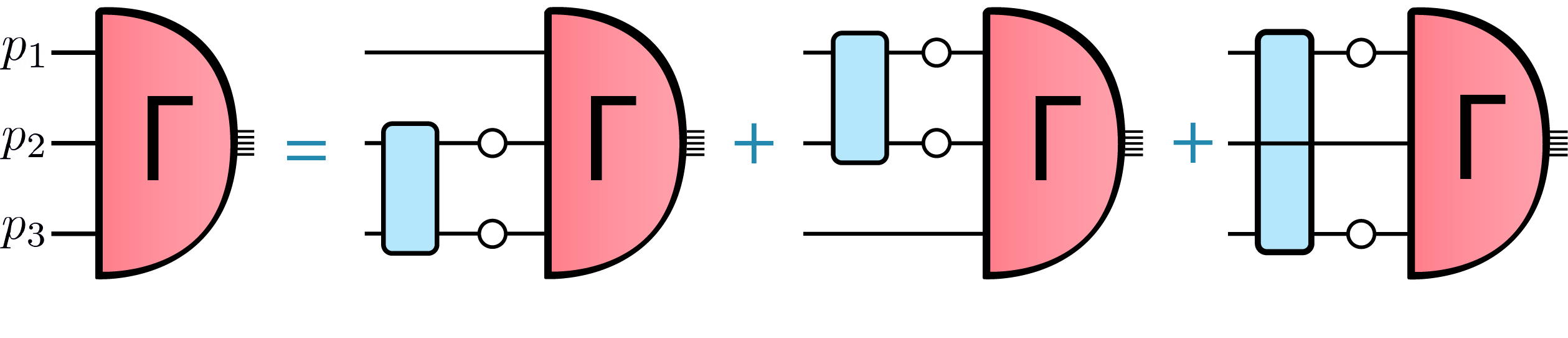}
   \caption{Three-body BSE~\eqref{Eq. FaddevBSE} with two-body kernels.}
   \label{Fig: Three-body} 
   \vspace{2mm}
   \includegraphics[width=0.9\columnwidth]{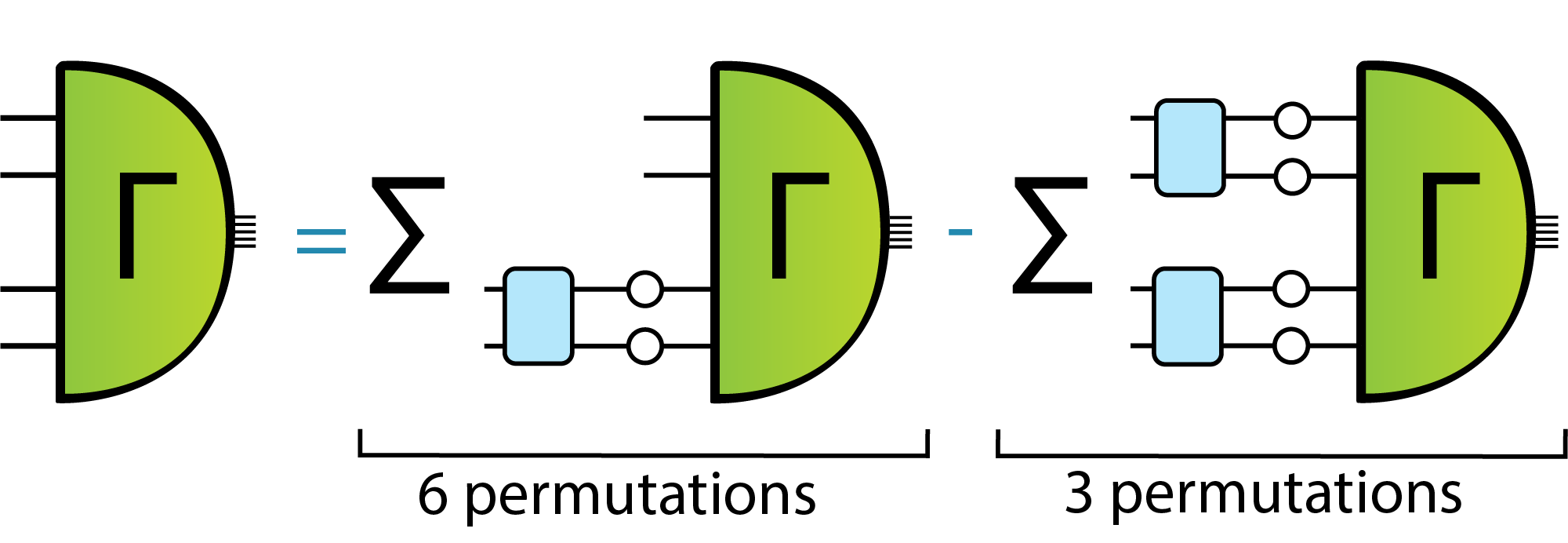}
   \caption{Four-body BSE \eqref{Eq. Four-body BSE } with two-body kernels and their counterterms.}
   \label{Fig: Four-body} 
 \end{figure}

The Bethe-Salpeter amplitude $\Gamma(\{p_i\})$ for a three-body system shown in Fig.~\ref{Fig: Three-body} 
depends on three momenta $p_1,p_2,p_3$, whose sum is the total onshell momentum $P$ with $P^2=-M_B^2$. The 
three-body BSE is the covariant Faddeev equation and can be written as 
\begin{equation}\label{Eq. FaddevBSE}
   \begin{split}
     \Gamma(\{p_i\}) &=\sum^{3}_{a}\Gamma_{(a)}(\{p_i\}) \,. \\
      \Gamma_{(a)}(\{p_i\})  &= \int \!\!\frac{d^4r}{(2\pi)^4}\, K(p_{a_1},q_{a_1},p_{a_2},q_{a_2}) \\
                              & \quad \times D(q_{a_1})\,D(q_{a_2})\,\Gamma(\{p_i\},a)\,.
   \end{split}
\end{equation}
In this case there are three possible two-body kernels
$K_{a}\in \{ K_{12},K_{13},K_{23} \}$,
whose sum does not lead to overcounting in 
the three-body scattering matrix $T^{(3)}$.
The amplitudes $\Gamma(\{p_i\},a)$ inside the loop follow from $\Gamma(\{p_i\})$ with the replacements
$p_{a_1}\to q_{a_1}$ and $p_{a_2}\to q_{a_2}$.

In practice it is useful to employ the solution strategy from Appendix C of Ref.~\cite{Eichmann:2011vu}, which 
takes advantage of the permutation-group properties of the amplitude and  reduces the numerical effort considerably. 
Here only one of the diagrams in Fig.~\ref{Fig: Three-body} needs to be calculated explicitly and the remaining ones are obtained by permutations,
\begin{equation}
  \Gamma(\{p_i\})=\Gamma_{(12)}(\{p_i\})+\Gamma_{(12)}(\{p'_i\})+\Gamma_{(12)}(\{p''_i\}),
\end{equation}
where $p'_i$ and $p''_i$ are the respective permuted momenta.
Instead of the three particle momenta $p_i$,
it is also useful to work with the relative momenta $q$ and $p$: 
\begin{alignat}{4}
	q&=\frac{p_2-p_1}{2}, \quad & p_1&=-q-\frac{p}{2}+\frac{1-\eta}{2}P, \nonumber \\ 
	p&=(1-\eta)\,p_3-\eta \,(p_1+p_2),  \quad & p_2&=q-\frac{p}{2}+\frac{1-\eta}{2}P, \nonumber \\ 
	P&=p_1+p_2+p_3,  \quad & p_3&=p+\eta P\,,   
\end{alignat}
with $\eta=1/3$ for equal masses. The resulting permuted relative momenta read 
\begin{alignat}{4}
   p'=&-q-\frac{p}{2}, \qquad & p'' &=q-\frac{p}{2}, \\ 
   q'=&-\frac{q}{2}+\frac{3p}{4}, \qquad & q''&=-\frac{q}{2}-\frac{3p}{4}\,.
\end{alignat}
We then solve the BSE amplitude for the variables $p^2$, $q^2$, $p\cdot q$, $p\cdot P$ and $q\cdot P$, which we
expand in Chebyshev and Legendre polynomials. 
This is what we refer to as `full solution' in the main text.

Like in the five-body system, we furthermore employ a singlet $\times$ pole approximation
\begin{equation}
   \Gamma(q,p,P)\approx f(\mathcal{S}_0)\sum_{a}\mathcal{P}_a\,.
   \label{Eq. 3-Body Pole Ansatz}
\end{equation}
Here the two-body pole for $a=12$ is  given by
\begin{equation}
  \mathcal{P}_{12}=\frac{1}{(p_1+p_2)^2+M^2_M} \,,
\end{equation}
with $M_M$ the mass of the two-body subsystem or `diquark',
and the singlet variable $\mathcal{S}_0$ is
\begin{equation}
   \mathcal{S}_0=\frac{q^2}{3}+\frac{p^2}{4} \,.
\end{equation}
Finally, in the singlet approximation we reduce the momentum dependence to $\mS_0$ by setting $\Gamma(q,p,P)\approx f(\mathcal{S}_0)$.

\subsection{\label{App. Four-body system} Four-body system}

In the four-body BSE shown in Fig.~\ref{Fig: Four-body}, one encounters the same situation discussed in Sec.~\ref{sec:subtr}:  
A naive summation of two-body kernels leads to overcounting, and   one 
needs subtraction terms~\cite{Huang1975,Kvinikhidze1992,Heupel:2012ua}. In this case there are six possible two-body kernels 
\begin{equation}\label{Eq. Four-Body 2Kerns}
   \begin{split}
      K_{a} \in \big\{ K_{12}\,, \; K_{13}\,, \; K_{14}\,, \; K_{23}\,, \; K_{24}\,, \; K_{34} \big\}
   \end{split}
\end{equation}
and three independent double-kernel configurations of the form 
\begin{equation}\label{Eq. Four-Body double 2Kerns}
  \begin{split}
     K_a\,K_b\in\big\{ K_{12}\,K_{34}\,, \; K_{13}\,K_{24}\,, \; K_{14}\,K_{23}  \big\}\,.
  \end{split}
\end{equation}
The four-body Bethe-Salpeter amplitude $\Gamma(\{p_i\})$ depends on four-momenta $p_1 \dots p_4$, whose sum is the total onshell
momentum $P$ with $P^2=-M_T^2$ ($T$ for `tetra'). The four-body equation is the  Faddeev-Yakubowski equation and  can be written as 
\begin{equation} \label{Eq. Four-body BSE }
   \begin{split}
      \Gamma(\{p_i\}) &= \sum_{a}^{6} \Gamma_{(a)}(\{p_i\}) - \sum_{a\neq b}^{3} \Gamma_{(a,b)}(\{p_i\})\,, \\
      \Gamma_{(a)}(\{p_i\})  &= \int \!\!\frac{d^4r}{(2\pi)^4}\, K(p_{a_1},q_{a_1},p_{a_2},q_{a_2}) \\
                              & \quad \times D(q_{a_1})\,D(q_{a_2})\,\Gamma(\{p_i\},a)\,, \\
      \Gamma_{(a,b)}(\{p_i\}) &= \int \!\!\frac{d^4r}{(2\pi)^4}\,  K(p_{a_1},q_{a_1},p_{a_2},q_{a_2}) \\
                              & \quad \times  D(q_{a_1})\,D(q_{a_2})\,\Gamma_{(b)}(\{p_i\},a)\,,
   \end{split}
\end{equation}
where the amplitudes $\Gamma(\{p_i\},a)$ inside the loop follow from $\Gamma(\{p_i\})$ with the replacements
$p_{a_1}\to q_{a_1}$ and $p_{a_2}\to q_{a_2}$.
In practice it is useful to work with the total momentum $P$ and the 
three relative momenta $p,q,k$ instead of the particle momenta $p_i$ with $i=1 \dots 4$. To this end, we employ the Jacobi momenta 
\begin{equation}
   \begin{split}
     p_1 &= \frac{k+q-p}{2}+\frac{P}{4}\,,\quad  p_3 = \frac{-k+p+q}{2}+\frac{P}{4}\,, \\ 
     p_2 &= \frac{k-q+p}{2}+\frac{P}{4}\,,\quad  p_4 = -\frac{k+p+q}{2}+\frac{P}{4}\,. 
   \end{split}
\end{equation}
Applying the same solution strategy as in the three-body case discussed above, 
we relate all amplitudes on the r.h.s. of Fig.~\ref{Fig: Four-body} to one diagram through permutations.
This also applies to the double-kernel diagrams, which can be obtained by multiplying the kernel onto the respective amplitude once more.

In the four-body case, a full solution without any restrictions on the amplitude 
is numerically not feasible, so we employ the singlet $\times$ pole approximation~\cite{Wallbott:2019dng,Wallbott:2020jzh,Hoffer:2024alv}
\begin{equation}
   \Gamma(q,p,k,P)\approx f(\mathcal{S}_0)\sum_{a}\mathcal{P}_{aa'}\,,
\end{equation}
where the two-body poles for $aa' = (12)(34)$ are given by 
\begin{equation}
   \mathcal{P}_{(12)(34)}=\frac{1}{(p_1+p_2)^2+M^2_M}\frac{1}{(p_3+p_4)^2+M^2_M}
\end{equation}
and $M_{M}$ is the mass of the two-body subsystem. The singlet variable $\mS_0$ is
\begin{equation}
   \mathcal{S}_0=\frac{k^2+q^2+p^2}{4}\,.
\end{equation}
In the singlet approximation, we further reduce the momentum dependence  to $\mS_0$ via $\Gamma(q,p,k,P)\approx f(\mathcal{S}_0)$.

\subsection{\label{App. Five Body} Five-body system}

The five-body BSE has already been discussed in the main text, Eqs.~(\ref{5bse-1}--\ref{S0}).
Its solution proceeds along the same lines as for the two- and three-body BSEs.
We employ the singlet $\times$ pole approximation~\eqref{Eq. 5Body - Pole Ansatz} 
and relate all terms in the BSE to one diagram such that 
\begin{equation}
  \Gamma_{(a)}(\{p_i\}) = \Gamma_{(12)}(\{p_i^{(a)}\})\,,
\end{equation}
where the $p_i^{(a)}$ are the permuted four-momenta.
 
 Furthermore, as discussed in Appendix D of Ref.~\cite{Eichmann:2025etg},
 for a five-body system one encounters for the first time a dimensional constraint relating the Lorentz invariants in Eq.~\eqref{li-variables}.
 Therefore, a singlet approximation where all variables except $\mS_0$ are set to zero is, strictly speaking, not possible
 because in that case also $\mS_0$ would vanish. To this end, we keep one of the variables $\eta_i$ from Eq.~\eqref{li-variables}
 in the system but set it to a fixed value, so that the amplitude still depends only on $\mS_0$.
 The singlet approximation in the five-body case is then analogous to the three- and four-body systems.

\bibliography{bib}%

\end{document}